\documentclass[aps,showpacs,eqsecnum,twocolumn,superscriptaddress]{revtex4}

\usepackage{amsmath}
\usepackage{latexsym}
\usepackage{graphicx}
\usepackage{times}

\begin{document}

\title{A hybrid approach to black hole perturbations from
  extended matter sources}

\author{Valeria Ferrari} 
\affiliation{Dipartimento di Fisica,
Universit\` a di Roma ``La Sapienza" and %Sezione 
INFN
%ROMA1, piazzale Aldo Moro 2, I-00185 
Rome, Italy}

\author{Leonardo Gualtieri} 
\affiliation{Dipartimento di Fisica,
Universit\` a di Roma ``La Sapienza" and %Sezione 
INFN
%ROMA1, piazzale Aldo Moro 2, I-00185 
Rome, Italy}

\author{Luciano Rezzolla}
\affiliation{Max-Planck-Institut f\"ur Gravitationsphysik,
Albert-Einstein-Institut, Potsdam
%% 14476 
%% Golm,
Germany}
\affiliation{SISSA, International School for
Advanced Studies and %Sezione 
INFN 
%TRIESTE, Via Beirut 2, 34014 
Trieste, Italy}
\affiliation{Department of Physics, Louisiana State University, Baton
Rouge, Lousiana % LA 70803 
USA}
 
\begin{abstract} 
We present a new method for the calculation of black hole perturbations
induced by extended sources in which the solution of the nonlinear
hydrodynamics equations is coupled to a perturbative method based on
Regge-Wheeler/Zerilli and Bardeen-Press-Teukolsky equations when these
are solved in the frequency domain. In contrast to alternative methods in
the time domain which may be unstable for rotating black-hole spacetimes,
this approach is expected to be stable as long as an accurate evolution
of the matter sources is possible. Hence, it could be used under generic
conditions and also with sources coming from three-dimensional numerical
relativity codes. As an application of this method we compute the
gravitational radiation from an oscillating high-density torus orbiting
around a Schwarzschild black hole and show that our method is remarkably
accurate, capturing both the basic quadrupolar emission of the torus and
the excited emission of the black hole.
\end{abstract} 

%\date{\today}


\maketitle

%%%%%%%%%%%%%%%%%%%%%%%%%%%%%%%%%%%%%%%%%%%%%%%%%%%%%%%%%%%%%%%%%%%%%%%%%%% 
\section{Introduction}
\label{intro}
%%%%%%%%%%%%%%%%%%%%%%%%%%%%%%%%%%%%%%%%%%%%%%%%%%%%%%%%%%%%%%%%%%%%%%%%%%% 

It is well-known that matter falling into a black hole, or orbiting
around it, is a source of gravitational radiation.  The detection of
these waves, and of the imprint they carry on the excitation of the black
hole quasi normal modes (QNM), would provide an invaluable test of
General Relativity in the strong-field regime.  This process can be
studied using the theory of black hole perturbations, the origin of which
dates back to 1957 when Regge and Wheeler~\cite{RW} showed that the axial
perturbations of a Schwarzschild black hole can be studied by solving a
radial equation in the frequency domain, the angular part of the
perturbation being known in terms of a suitable expansion in tensor
spherical harmonics.

Among the many dedicated to this subject, two papers can be considered as
milestones in the theory of black hole perturbations. The first one is by
F. Zerilli~\cite{Zer}, who derived the radial equation governing the
polar perturbations of a Schwarzschild black hole, thus completing the
work of Regge and Wheeler (see~\cite{nr05} for a recent review). The
second one is by S. Teukolsky~\cite{Teu0}, who achieved the formidable
task of reducing the equations for the perturbations of a Kerr black hole
to a master radial equation.  Both the Zerilli and the Teukolsky
equations were derived in the frequency domain.

The theory of black hole perturbations has been developed largely to
study the stability properties of black holes and to determine the
gravitational signals emitted by point particles moving near black holes
on different orbits (see for instance~\cite{NOK} and references therein).
In addition, it has been used to evaluate back-reaction effects on the
orbital motion of a point particle moving around a nonrotating black
hole~\cite{CKP}.

Although it clearly represents only a toy-model, the point-particle
approximation has been very useful to study a number of
phenomena that occur in the neighborhood of black holes and it has also
been applied to study stellar perturbations~\cite{noi1,noi2}. However, if
one wants to investigate realistic astrophysical processes, in which
finite-size effects and internal dynamics need to be properly taken
into account, the point-particle approximation is certainly too crude.
Therefore, several attempts have been made towards a more realistic
description. In ref.~\cite{DS}, for instance, the point particle orbiting
a black hole was treated as having a spin angular momentum, while in
ref.~\cite{HSW} the infalling matter was assumed to be a finite size
sphere of dust. Furthermore, in ref.~\cite{SW} a semianalytic method was
developed to study the radial infall of a shell of dust; while this
latter approach has the advantage of being easily generalizable to
sources with arbitrary shapes, it has the serious drawback that it can be
applied only to sources in which the matter configurations are freely falling
onto the black hole. It is worth stressing that in all cases mentioned
above the black hole is assumed to be nonrotating.

Alternative and more general approaches to treat extended sources have
recently been developed by Papadopulous and Font~\cite{PF99} and by Nagar
et al.~\cite{NFZD}, by combining perturbative approaches with fully
multidimensional hydrodynamical calculations. In ref.~\cite{NFZD}, in
particular, the polar perturbations of a Schwarzschild black hole induced
by an oscillating, thick accretion disk (a torus) have been studied by
solving the inhomogenous Zerilli equation in the {\it time domain}. The
source terms were computed self-consistently by using the results of
two-dimensional (2D) simulations in which the relativistic hydrodynamical
equations were solved for a torus oscillating in the Schwarzschild,
unperturbed background. The purpose of that work was to compare the
gravitational signal found by solving the Zerilli equation, with the one
found using the standard quadrupole formalism applied to the oscillating
torus.

In this paper we develop a new approach to study the perturbations of a
Schwarzschild black hole excited by extended matter sources, by solving
the inhomogeneous Bardeen-Press-Teukolsky (BPT) equation~\cite{BP,Teu} in
the {\it frequency domain}. Also in this case the source for the
perturbative equations is the result of independent 2D hydrodynamical
simulations in the time domain. Because of the combined use of both time
and frequency domains approaches, we refer to this as to a {\it hybrid
approach}.

There are at least three good reasons why our hybrid approach may be
preferrable over a pure time-domain approach as the one used
in~\cite{NFZD}. Firstly, the Regge-Wheeler and the Zerilli equations
(hereafter RWZ equations) cannot be generalized to rotating black holes,
whereas the BPT equation admits such a generalization, {\it i.e.,} the
Teukolsky equation (We recall that the Teukolsky equation is not
separable in the time domain because the eigenvalues of the angular
functions are frequency dependent; this is not the case in the frequency
domain where the equation is indeed separable.). Secondly, although the
BPT equation is intrinsically unstable when integrated in the time
domain, its integration does not offer any numerical difficulty when
performed in the frequency domain. Thirdly, it is a common experience
that by removing the error in the time integration of a system
oscillating around its equilibrium state, a frequency-domain approach
improves the overall accuracy. The work presented here serves therefore
to develop a new procedure for the solution of the BPT equation with
extended matter sources, whose generalization will allow to reach a
long-standing goal: {\it i.e.,} the study of the gravitational-wave
emission from an extended source perturbing a {\em rotating} black hole.

While the mathematical apparatus behind our hybrid approach is well-know,
our numerical implementation is more innovative and to test it against an
astrophysically realistic and computationally complex configuration we
have here also considered the gravitational-wave emission from a
high-density torus oscillating in its orbital motion around a
Schwarzschild black hole. This is the same source considered in
ref.~\cite{NFZD} and thus offers the possibility of comparing for the
same test case our hybrid approach with one which is instead in the
time-domain only.

To this scope, we recall that the possibility that extremely dense and
massive accretion tori form around a black hole has been considered in
recent years, initially as a possible explanation of $\gamma$-ray
bursts~\cite{Woo}.  In these systems, a torus of mass $M_t\sim
0.1\!-\!1M_\odot$ and mean density as high as $\sim10^{11}\!-\!10^{12}$
g/cm$^3$, forms around a black hole of a few solar masses. The torus can
be located within a few tens of km from the horizon and could also be
subject to a dynamical instability which would induce its accretion onto
the black hole on a dynamical timescale (see~\cite{font:02,ZRF} and
references therein). High-density tori could be generated in many
astrophysical scenarios (see~\cite{Woo,ZRF} for a general overview): in
the coalescence of black hole-neutron star or neutron star binaries, as
shown by several numerical simulations~\cite{st06} in the collapse to a
black hole of a rapidly rotating supramassive neutron
star~\cite{VS,detal_06}, as a byproduct of collapsars, that is, massive
stars in which iron core-collapse does not produce an explosion, but
forms a black hole~\cite{MW}.  As discussed in~\cite{ZRF}, these systems
can be created with event rates comparable to that of core collapse
supernovae. In addition, because the tori have an intrinsically large
quadrupole moment, even small-amplitude oscillations would produce
gravitational signals that may be detectable by ground based
interferometers Virgo and LIGO~\cite{ZRF}. As a result, the calculation
of the gravitational waves from these tori does not only represent a
useful testbed for our novel approach, but it also offers the opportunity
for a more accurate calculation of the emission from a realistic and
intense source of gravitational radiation.

The gravitational radiation emitted by a torus oscillating around a black
hole was first computed in ref.~\cite{ZRF} within the Newtonian
quadrupole formalism.  However, the quadrupole formalism may not be able
to describe a system with such a high energy density, and located so
close to a black hole horizon with sufficient accuracy. In particular,
the possible excitation of the black hole QNM cannot be revealed by the
quadrupole approach.  The same system has been studied in~\cite{NFZD},
where the perturbed equations have been integrated in the time domain and
no excitation of the black hole quasi-normal modes (QNMs) was
found. Conversely, as we will later discuss, our hybrid method applied to
the same system allows to catch the excitation of the black hole quasi
normal modes, though the peak they produce in the energy spectrum is much
smaller than that corresponding to the torus oscillation.

The plan of the paper is as follows: in Section~\ref{approach} we review
the main equations which describe the perturbations of a Schwarzschild
black hole in the RWZ and in the BPT approaches when these are expressed
in the frequency domain. Furthermore, we also recall the expression of
the gravitational energy flux computed within the Newtonian quadrupole
formalism. In Section~\ref{hybr} we present the analytical and numerical
setup of our approach for the solution of both the perturbative and
general relativistic hydrodynamics equations. Finally, in
Section~\ref{results} we discuss the results of the numerical
integrations for the representative case of an oscillating torus, while
in Section~\ref{conclusions} we draw our conclusions. A number of Appendices
contain details on the form of the source functions employed for the
solution of the perturbative equations.

%%%%%%%%%%%%%%%%%%%%%%%%%%%%%%%%%%%%%%%%%%%%%%%%%%%%%%%%%%%%%%%%%%%%%%%%%%% 
\section{A brief review of black hole perturbations in the frequency domain}
\label{approach}
%%%%%%%%%%%%%%%%%%%%%%%%%%%%%%%%%%%%%%%%%%%%%%%%%%%%%%%%%%%%%%%%%%%%%%%%%%% 

In what follows we briefly summarize the relevant equations needed to
describe the perturbations of a Schwarzschild spacetime in the RWZ and in
the BPT formulations, as they will later be used in our hybrid approach.
We will also write the expression of the energy flux computed within the
Newtonian quadrupole approximation since it will be used for comparison
with the corresponding quantities computed within the perturbative
approaches.

%--------------------------------------------------------------------------
\subsection{The RWZ approach}
%--------------------------------------------------------------------------

In the RWZ approach~\cite{RW,Zer}, the equations describing the radial
behaviour of the perturbations of a Schwarzschild black hole are reduced
to two wave equations after expanding all perturbed tensors in tensorial
spherical harmonics and after separating the radial from the angular
part. In particular, they are expressed for two suitable combinations of
the components of the perturbed metric tensor and which are referred to
as the Zerilli function, ${\widetilde Z}^{(+)}_{\ell m}$, and the
Regge-Wheeler function, ${\widetilde Z}^{(-)}_{\ell m}$,
respectively. Hereafter we will indicate with a ``tilde'' all the
perturbative variables that have a time dependence which is not
explicitly indicated and with $\ell,m$ the indices of the spherical
harmonic decomposition.

Since the system we will consider, namely the oscillating torus perturbing the
spherically symmetric black hole spacetime, is axially symmetric, the
perturbations are also axially symmetric, and
hence in what follows we will consider $m=0$, with the Zerilli and
Regge-Wheeler functions expressed as ${\widetilde Z}^{(+)}_{\ell}(r,t)$,
${\widetilde Z}^{(-)}_{\ell}(r,t)$, with $\ell\ge 2$.

Their Fourier transforms are simply given by  
\begin{equation}
Z^{(\pm)}_{\ell}(r,\omega) \equiv \frac{1}{2\pi}\int e^{{\rm i}\omega t}
{\widetilde Z}^{(\pm)}_{\ell}(r,t) dt \ ,
\end{equation}
and satisfy the inhomogeneous wave equations
\begin{eqnarray}
{\boldsymbol {\rm I\!\!\,L}}^{(\pm)}Z^{(\pm)}_{\ell}=
	S_{\ell}^{(\pm)}\ ,
\label{ZerRWeq}
\end{eqnarray}
where the upper index ``$+$'' refers to the Zerilli equation, the ``$-$''
one to the Regge-Wheeler equation, and the symbol ${\boldsymbol {\rm
    I\!\!\,L}}^{(\pm)}$ is just a shorthand for the differential operator
\begin{equation}
{\boldsymbol {\rm I\!\!\,L}}^{(\pm)} \equiv
\partial_{r_*}^2+\omega^2-V^{(\pm)}\ .
\label{waveoperator}
\end{equation}
Note that in order to provide mathematically consistent boundary
conditions at the event horizon, the latter is moved to negative spatial
infinity in terms of the ``tortoise'' radial coordinate \hbox{$r_*\equiv
  r+2M\ln\left({r}/{2M}-1\right)$}, and that the Zerilli and
Regge-Wheeler potentials in equation~\eqref{ZerRWeq} have explicit
expressions given respectively by
\begin{eqnarray}
&& \hskip -1.0 cm
V^{(+)}\equiv\frac{2(r-2M)}{r^4(\Lambda r+3M)^2}
	\bigl[\Lambda^2(\Lambda+1)r^3
	+ \nonumber\\ &&\hskip 1.5 cm
+3M\Lambda^2r^2+9M^2\Lambda r+9M^2\bigr]\ ,\\ 
&&\hskip -1.0 cm
	V^{(-)}\equiv\frac{2(r-2M)}{r^4}\left[(\Lambda+1)r-3M\right]\ .
\label{RWpot}
\end{eqnarray}
Here $\Lambda\equiv(\ell-1)(\ell+2)/2$, while
$S^{(\pm)}_{\ell}(r,\omega)$ are suitable combinations of the components
of the stress-energy tensor of the extended matter source, after having
been expanded in tensor spherical harmonics and Fourier transformed in
time. Explicit expressions for the source functions
$S^{(\pm)}_{\ell}(r,\omega)$ can be found in Appendix~\ref{sources_rwz}.

Equations (\ref{ZerRWeq}) can to be solved numerically by imposing
outgoing boundary conditions at spatial infinity ({\it i.e.,} $r, r_* \to
\infty$), and ingoing boundary conditions at the event horizon ({\it
  i.e.,} $r = 2M,~r_* \to -\infty$). The energy spectrum of the
gravitational radiation measured by a distant observer and integrated over the
solid angle, can then be expressed in terms of the asymptotic amplitude
of $Z^{(+)}_{\ell}$ and $Z^{(-)}_{\ell}$ as
\begin{eqnarray}
&& \hskip -1.1 cm
\frac{dE^{^{\rm ZRW}}}{d\omega}=
\sum_{l=2}^\infty\left.\frac{dE_{\ell}}{d\omega}\right|_{r,r_* \to  \infty} 
\nonumber \\
&&\hskip 0.2 cm
=\sum_{l=2}^\infty
\frac{1}{8}\frac{(\ell+2)!}{(\ell-2)!}\left[\omega^2\left(Z^{(+)}_{\ell}\right)^2
+\left(Z^{(-)}_{\ell}\right)^2
\right]\,.
\label{dEdw1}
\end{eqnarray}

%--------------------------------------------------------------------------
\subsection{The BPT approach}
%--------------------------------------------------------------------------

In the BPT approach~\cite{BP,Teu}, the perturbations of the Weyl scalar
$\delta \Psi_4$, which describes the outgoing gravitational radiation,
are expanded in spin-weighted spherical harmonics $_{-2}S^*_{\ell m}$ and
integrated over the solid angle $d\Omega = \sin\theta d\theta d\phi$
\begin{equation}
{\widetilde {\Psi}}_{\ell m}(r,t)=\int d\Omega~_{-2}S^*_{\ell m}(\vartheta,\phi) 
r^4 \delta {\widetilde \Psi}_4(t,r,\vartheta,\phi)\ .
\end{equation}
Also in this case, being the perturbing stress-energy tensor axially symmetric
the only nonvanishing components are those with $m=0$ and
hence perturbed Einstein equations take the form of inhomogeneous wave
equations 
\begin{equation}
{\boldsymbol {\rm I\!\!\ L}}^{^{\rm BPT}}{\Psi}_{\ell}(r,\omega)=
-S^{^{\rm BPT}}_{\ell}(r,\omega)\ ,
\label{bpt}
\end{equation}
where $\Psi_{\ell}(r,\omega)$ is the Fourier transform of ${\widetilde
  \Psi}_{\ell}(r,t)$, the symbol ${\boldsymbol {\rm I\!\!\,L}}^{^{\rm
    BPT}}$ is just a shorthand for the differential operator,
\begin{equation}
{\boldsymbol {\rm I\!\!\ L}}^{^{\rm BPT}}\equiv 
\Delta^2\partial_r\frac{1}{\Delta}\partial_r+U(r)\ , 
\end{equation}
with $\Delta\equiv r(r-2M)$ and the potential $U$ having explicit form
\begin{equation}
U\equiv\frac{r^4\omega^2+4{\rm i}(r-M)r^2\omega}{\Delta}
-8{\rm i}\omega r-2\Lambda\ .
\end{equation}
As for the RWZ equation, the source terms $S^{^{\rm
    BPT}}_{\ell}(r,\omega)$ are suitable combinations of components of
the stress-energy tensor after having been expanded in spin-weighted
spherical harmonics and Fourier transformed. Explicit expressions for the
source functions $S^{(\pm)}_{\ell}(r,\omega)$ can be found in
Appendix~\ref{sources_bpt}.

Equation (\ref{bpt}) can be integrated numerically imposing the same
boundary conditions discussed for the RWZ equation. In this case, the
energy spectrum of the gravitational radiation collected by a distant
observer and integrated over the solid angle is expressed as
\begin{equation}
\frac{dE^{^{\rm BPT}}}{d\omega}=
\sum_{\ell=2}^\infty\left.\frac{dE_{\ell}}{d\omega}\right|_{r_* \to \infty}=
\sum_{\ell=2}^\infty\frac{8\pi^2}{\omega^2}\left|\frac{\Psi_{\ell}}{r^3}\right|^2\ .
\label{dEdw2}
\end{equation}

%--------------------------------------------------------------------------
\subsection{The Newtonian quadrupole approximation}
%--------------------------------------------------------------------------

The Newtonian quadrupole approximation is based on the assumptions that
the gravitational field is weak, that the velocity of matter in the
source is much smaller than the speed of light, and that the source
itself is small compared to the characteristic wavelength of the emitted
gravitational radiation. In this approach, then, the amplitude of the
gravitational wave is estimated in terms of the quadrupole tensor
associated to the source
\begin{equation}
{\tilde q}_{ij}\equiv\int_V{\widetilde T}^{00}(t,\vec x)
	x^ix^jd^3x~~~~~(i,j=1,2,3)\ .
\label{qdef}
\end{equation}
where ${\boldsymbol T}$ is the stress-energy tensor of the matter
  source. The resulting gravitational-wave energy-spectrum is then given
  by 
\begin{equation}
\frac{dE^{\rm quad}}{d\omega}=\frac{4\pi}{5}
\omega^6\sum_{ij}\left|Q_{ij}(\omega) Q_{ij}(\omega)\right|
\label{dEdw3}
\end{equation}
where $Q_{ij}(\omega)$ is the Fourier transform of the traceless
quadrupole tensor ${\widetilde Q}_{ij}$
\begin{equation}
{\widetilde Q}_{ij}\equiv {\tilde q}_{ij}-\frac{1}{3}\delta_{ij}
\sum_{kl}\delta_{kl}{\tilde q}_{kl}\ .\label{qtr}
\end{equation}

As we will comment more extensively in Section~\ref{results}, we will
compare expressions \eqref{dEdw1}, \eqref{dEdw2} and \eqref{dEdw3} for
the gravitational emission coming from an oscillating torus and point out
the respective advantages and disadvantages.

%%%%%%%%%%%%%%%%%%%%%%%%%%%%%%%%%%%%%%%%%%%%%%%%%%%%%%%%%%%%%%%%%%%%%%%%%%% 
\section{The Hybrid Approach}
\label{hybr}
%%%%%%%%%%%%%%%%%%%%%%%%%%%%%%%%%%%%%%%%%%%%%%%%%%%%%%%%%%%%%%%%%%%%%%%%%%% 

We will now give a general and brief account of our hybrid approach,
sketching its most relevant parts and its numerical implementation.  We
will first discuss how to construct the components of the stress-energy
tensor, needed as source terms in the perturbative equations, through the
solution of the 2D hydrodynamic equations in a Schwarzschild background.
We will then illustrate how to find the solution of the RWZ and BPT
equations once the sources are known in the frequency domain.

%-------------------------------------------------------------------
\subsection{Computing the sources from the hydrodynamic equations}
\label{hydrosource}
%-------------------------------------------------------------------

As mentioned in the Introduction, any realistic and extended matter
source whose dynamics goes beyond the elementary free-fall from infinity,
will need to be described through the solution of the general
relativistic hydrodynamics equations on the background black hole
spacetime. In particular, this is accomplished by computing at the
different spacetime points covered by a numerical grid, the components of
the stress-energy tensor $T_{\mu\nu}(t,r,\theta)$ necessary for the
calculation of the sources ${\widetilde S}^{(\pm)}_{\ell}$ and
${\widetilde S}^{^{\rm BPT}}_{\ell}$.

For a matter source represented by a perfect fluid with four-velocity
${\boldsymbol u}$ and described by the stress-energy tensor
\begin{equation}
\label{stress-tensor}
T_{\mu\nu}\equiv (e+p)u_\mu u_\nu+p g_{\mu\nu}
        = \rho h u_\mu u_\nu+p g_{\mu\nu} \ ,
\end{equation}
where $g_{\mu\nu}$ are the coefficients of the metric which we choose to
be Schwarzschild. Here, $e$, $p$, $\rho$, and $h = (e+p)/\rho$ are the
proper energy density, the isotropic pressure, the rest-mass density, and
the specific enthalpy, respectively. In practice, all of these quantities
are computed by the solution of the conservation equations for
the stress-energy tensor and baryon number density
\begin{eqnarray}
\label{cset}
&& \nabla_{\mu} T^{\mu\nu} = 0\ , \\
\label{cbnd}
&& \nabla_{\mu} (\rho u^{\mu}) = 0 \ , 
\end{eqnarray}
where $\nabla$ indicates the covariant derivative in the background
Schwarzschild spacetime, together with an equation of state (EOS)
relating the pressure to other thermodynamical quantities. For simplicity
we will hereafter model the fluid as ideal with a polytropic $p=\kappa
\rho^\gamma=\rho\epsilon(\gamma-1)$, where $\epsilon=e/\rho - 1$ is the
specific internal energy, $\kappa$ is the polytropic constant and
$\gamma$ is the adiabatic index.

In order to preserve their conservative nature, we cast
equations~\eqref{cset},~\eqref{cbnd} in the form of a flux-conservative
hyperbolic system after introducing suitable {\it ``conserved''}
variables rather than in terms of the ordinary fluid, or {\it
  ``primitive''}, variables. In this case, equations
\eqref{cset}-\eqref{cset} assume the form~\citep{baiotti04}
\vbox{
\begin{eqnarray}
\label{fcf}
&&\hskip -1.5cm \frac{\partial {\bf U}({\bf w})}{\partial t} +
	\frac{\partial [\sqrt{-g_{00}}{\bf F}^{r}({\bf w})]}{\partial r} +
\nonumber \\ \nonumber \\ 
&&\hskip 0.6cm	\frac{\partial [\sqrt{-g_{00}}{\bf F}^{\theta}({\bf w})]}
	     {\partial \theta}
	 = {\bf S}({\bf w}) \ ,
\label{system}
\end{eqnarray}}
where ${\bf U}({\bf w})=(D, S_r, S_{\theta}, S_{\phi})$, ${\bf F}^{i}$
and ${\bf S}$ are the state-vector, the fluxes and the sources of the
evolved quantities, respectively (see~\cite{font:02} for the explicit
expressions for ${\bf F}^{i}$ and ${\bf S}$ in a Schwarzschild
spacetime). The following set of equations 
\begin{eqnarray}
\label{evolved}
D &\equiv&  \rho \Gamma \ ,
\nonumber \\
S_j &\equiv&  \rho h \Gamma^2 v_j \ ,
\end{eqnarray}
together with the ideal-fluid EOS provide the relation between the
conserved and primitive variables in the vector ${\bf w} = (\rho, v_{i},
\epsilon)$. Here $\Gamma\equiv\alpha u^t = (1-v^2)^{-1/2}$, where $v^2
\equiv \gamma_{ij}v^i v^j$ is the Lorentz factor measured by a local
static observer. Note that the covariant components of the three-velocity
are defined in terms of the spatial 3-metric $\gamma_{ij}$ to be
$v_i=\gamma_{ij}v^j$, where $v^i=u^i/\alpha u^t$. (Although in
axisymmetry, we evolve also the azimuthal component of the equations of
motion, so that the index $j$ takes the values $j=r,\theta,\phi$.)

	The numerical code used in our computations is the same used in
ref.~\cite{ZRF} and it performs the numerical integration of system
(\ref{system}) using upwind high-resolution shock-capturing (HRSC)
schemes based on approximate Riemann solvers. Exploiting the flux
conservative form of equations (\ref{fcf}), the time evolution of the
discretized data from a time-level $n$ to the subsequent one $n+1$ is
performed according to the following scheme
\vbox{
\begin{eqnarray}
\label{godunov}	
&& \hskip -1.0cm {\bf U}_{i,j}^{n+1} = {\bf U}_{i,j}^{n}
	-  \frac{\Delta t}{\Delta r}
	\left(\widehat{{\bf F}}^r_{i+1/2,j}-
	      \widehat{{\bf F}}^r_{i-1/2,j}\right) -
\nonumber \\ \nonumber \\ 
	&& \hskip 0.4cm \frac{\Delta t}{\Delta \theta}
    	\left(\widehat{{\bf F}}^{\theta}_{i,j+1/2}-
  	      \widehat{{\bf F}}^{\theta}_{i,j-1/2}\right) +
\Delta t \ \  {\bf S}_{i,j} \ ,
\end{eqnarray}}
\noindent
where the subscripts $i,j$ refer to spatial ($r,\theta$) grid points, so
that ${\bf U}_{i,j}^n \equiv {\bf U}(r_i, \theta_j, t^n)$.  The
inter-cell numerical fluxes, $\widehat{{\bf F}}^{r}_{i \pm 1/2,j}$ and
$\widehat{{\bf F}}^{\theta}_{i,j \pm 1/2}$, are computed using Marquina's
approximate Riemann solver~\cite{ZRF}. A piecewise-linear cell
reconstruction procedure provides second-order accuracy in space, while
the same order in time is obtained with a conservative two-step
second-order Runge-Kutta scheme applied to the above time update.

Our computational grid consists of $N_r \times N_{\theta} = 250 \times
84$ zones in the radial and angular directions, respectively, covering a
computational domain extending from $r_{\rm {min}}=2.1$ to $r_{\rm
  {max}}=30$ and from $0$ to $\pi$. The radial grid is logarithmically
spaced in terms of the tortoise coordinate with the maximum radial
resolution at the innermost grid being $\Delta r=6\times 10^{-4}$. As in
ref.~\cite{ZRF}, we use a finer angular grid in the regions that are
usually within the torus and a much coarser one outside. The boundary
conditions adopted, the treatment of the vacuum region outside the torus
with a low density atmosphere, and the procedure for recovering physical
variables from the conserved quantities $D$ and $S_{i}$ are the same as
those used in ref.~\cite{ZRF}. The interested reader is referred to that
work for further details.

%--------------------------------------------------------------------------
\subsection{Solution of the RWZ and BPT equations}
\label{RWZequation}
%--------------------------------------------------------------------------

Before solving eq.~(\ref{ZerRWeq}) with ingoing wave boundary condition
at the horizon and outgoing wave boundary condition at infinity, we
determine the solution of the corresponding homogeneous equation
\begin{equation}
{\boldsymbol {\rm I\!\!\ L}}^{(\pm)}Z_{\ell}^{(\pm)}=0
\label{ZRWhom}
\end{equation}
for an assigned value of $\omega$, finding two particular solutions,
$Z_{\ell}^{(\pm)~{\rm in}},\ Z_{\ell}^{(\pm)~{\rm out}}$, which satisfy,
respectively, ingoing wave boundary conditions at the horizon
\begin{equation}
Z_{\ell}^{(\pm)~{\rm in}}
(r_* \rightarrow -\infty,\omega)=e^{-{\rm i}\omega r_*},
\label{eqZin}
\end{equation}
and outgoing wave boundary conditions at infinity
\begin{equation}
Z_{\ell}^{(\pm)~{\rm out}}
(r_*\rightarrow\infty,\omega)=e^{{\rm i}\omega r_*}\ .
\label{eqZout}
\end{equation}
Since they are independent solutions, their Wronskian
\begin{equation}
W_*\equiv 
 Z_{\ell}^{(\pm)~{\rm out}} 
 \left(\frac{\partial Z_{\ell}^{(\pm)~{\rm in}}}{\partial r_*}\right)
 -  \left(\frac{\partial Z_{\ell}^{(\pm)~{\rm out}}}{\partial r_*}\right)
 Z_{\ell}^{(\pm)~{\rm in}}
\end{equation}
is constant. The solution of the inhomogeneous equations (\ref{ZerRWeq}),
$Z_{\ell}^{(\pm)}$, is then found as a convolution integral of
the source with $Z_{\ell}^{(\pm)~{\rm in}}$ and $Z_{\ell}^{(\pm)~{\rm
    out}}$
\begin{eqnarray}
Z_{\ell}^{(\pm)}(r,\omega)&=&
\frac{Z_{\ell}^{(\pm)~{\rm out}}}{W_*}\int_{2M}^{r}dr'
	\frac{Z_{\ell}^{(\pm)~{\rm in}}
          \ S_{\ell}^{(\pm)}\ r^{\prime 2}}{\Delta}
        \nonumber\\ &+&
        \frac{Z_{\ell}^{(\pm)~{\rm out}}}{W_*}
        \int_{r}^{\infty}dr' \frac{Z_{\ell}^{(\pm)~{\rm out}}
          \ S_{\ell}^{(\pm)}\ r^{\prime 2}}{\Delta} \ .
\nonumber\\
\label{gensolZRW}
\end{eqnarray}
At spatial infinity the Zerilli function can be written as \cite{DRPP}
\begin{equation}
Z_{\ell}^{(\pm)}(r,\omega)\rightarrow{\cal A}_{\ell}^{(\pm)}(\omega) 
	e^{{\rm i}\omega r}\ ,
\end{equation}
where
\begin{equation}
{\cal A}_{\ell}^{(\pm)}(\omega) \equiv \frac{1}{W_*}\int_{2M}^{\infty}dr
	\frac{Z_{\ell}^{(\pm)~{\rm in}}
          \ S_{\ell}^{(\pm)}\ r^{2}}{\Delta}\ .
\label{Hrwz}
\end{equation}
In this way, the functions ${\cal A}^{(+)}_{\ell}$ and ${\cal
  A}^{(-)}_{\ell}$ represent the amplitudes of the gravitational
radiation with polar and axial parity and, as expressed by~\eqref{dEdw1},
they both contribute to the gravitational spectrum
\begin{equation}
\frac{dE^{^{\rm RWZ}}}{d\omega}=\sum_{\ell=2}^\infty
	\frac{1}{8}\frac{(\ell+2)!}{(\ell-2)!}\left[\omega^2 
	\left({\cal A}_{\ell}^{(+)}\right)^2 +
        \left({\cal A}_{\ell}^{(-)}\right)^2 \right]\ .
\label{dEdw1bis}
\end{equation}

As done for the RWZ equations, also for the BPT equation (\ref{bpt}) we first 
solve the corresponding homogeneous equation
\begin{equation}
{\boldsymbol {\rm I\!\!\ L}}^{^{\rm BPT}}\Psi_{\ell}(r,\omega)=0
\label{BPThom}
\end{equation}
for an assigned value of $\omega$ and find two independent solutions
which satisfy, respectively, the condition of a pure ingoing wave at the
black hole horizon and of a pure outgoing wave at infinity
\begin{eqnarray}
\Psi_{\ell}^{\rm in}(r_* \rightarrow
-\infty,\omega)&=&\Delta^2e^{-{\rm i}\omega r_*} \ ,
\label{Psiin}\\
&&\nonumber\\
\Psi_{\ell}^{\rm out}(r_*\rightarrow\infty,\omega)&=&r^3e^{{\rm i}\omega r_*}\ ,
\label{Psiout}
\end{eqnarray}
where the correcting factors $\Delta$ and $r^3$ in~\eqref{Psiin} and
~\eqref{Psiout} are the result of the asymptotic expansion of the BPT
equation~\cite{BP,Teu}.

The solution $\Psi_{\ell}(r,\omega)$ of the inhomogeneous equation
(\ref{bpt}) has a form analogous to the one for the RWZ equations
(\ref{gensolZRW}) and also in this case for $r_*\rightarrow\infty$ it can
be expressed as
\begin{equation}
\Psi_{\ell}(r_*,\omega)\rightarrow {\cal A}_{\ell}^{^{\rm
    BPT}}(\omega)r^3e^{{\rm i}\omega r_*}\ ,
\end{equation}
with
\begin{equation}
{\cal A}_{\ell}^{^{\rm BPT}}(\omega)=-\frac{1}{W}\int_{2M}^{\infty}dr'
\frac{\Psi_{\ell}^{\rm in}\ S_{\ell}^{^{\rm
BPT}}(r,\omega)\ }{\Delta^2}
\label{Hbpt}
\end{equation}
where the Wronskian is now defined as
\begin{equation}
W\equiv \frac{1}{\Delta}
	\left[\Psi_{\ell}^{\rm in}
	\left(\frac{\partial \Psi_{\ell}^{\rm out}}{\partial r}\right)-
	\left(\frac{\partial \Psi_{\ell}^{\rm in}}{\partial r}\right)
	\Psi_{\ell}^{\rm out}\right]\ .
\end{equation}
Also in this case, the complex amplitude ${\cal A}^{^{\rm BPT}}_{\ell}$
describes the gravitational-wave amplitudes with polar and axial parity,
so that the gravitational spectrum (\ref{dEdw2}) becomes
\begin{equation}
\frac{dE^{^{\rm BPT}}}{d\omega}=
\sum_{\ell=2}^\infty\frac{8\pi^2}{\omega^2}\left|{\cal A}^{^{\rm
BPT}}_{\ell}\right|^2\ .
\label{dEdw2bis}
\end{equation}

It is important to note that the solutions $\Psi_{\ell}^{\rm in/out}$ of
the homogeneous equations (\ref{Psiin}) and (\ref{Psiout}), can be
expressed in terms of the solutions $Z^{(-)~{\rm in/out}}_{\ell}$ of the
corresponding homogeneous RWZ equations (\ref{ZRWhom}) through the {\em
  Chandrasekhar transformation}~\cite{chandrabook} 
\begin{eqnarray}
&& \hskip -0.2cm  \Psi_{\ell}^{\rm in/out}=
\frac{r^3}{8\omega}\sqrt{\frac{(\ell+2)!}{(\ell-2)!}}
\left[V^{(-)}Z^{(-)~{\rm in/out}}_{\ell}\right.\nonumber\\
&&\hskip -0.4cm 
\left.+ 2\left(\frac{r-3M}{r^2}+{\rm i}\omega\right)
\left(\frac{\Delta}{r^2}Z^{(-)~{\rm in/out}}_{\ell , r}+{\rm i} \omega
Z^{(-)~{\rm in/out}}_{\ell}\right)\right]\,.\nonumber\\
\label{chandra}
\end{eqnarray}
In what follows we will use this transformation to avoid the resolution
of~\eqref{BPThom}.

%--------------------------------------------------------------------------
\subsection{Combining the two approaches}
\label{implementation}
%--------------------------------------------------------------------------

Having described the distinct approaches for the calculation of the
sources in the time domain and the solution of the perturbative equations
in the frequency domain, we now discuss how to combine the two methods
within our hybrid approach. Since the methodology applies unchanged
whether we are considering the solution of the RWZ or of the BPT
equations, we will drop this distinction in what follows.

Assume therefore that the solution of the hydrodynamical
equations~\eqref{cset},~\eqref{cbnd} as illustrated in
Sect.~\ref{hydrosource} has provided the components of the stress-energy
tensor $T_{\mu\nu}(t,r,\theta)$ for all the spacetime events that are
relevant. The first step for the effective calculation of the sources
consists then in the removal of the angular dependence. This is done by
calculating for the different values of $\ell$, the integrals over
the 2-sphere of the stress-energy tensor components, as given in
eqs.~\eqref{intRWZ1}--\eqref{intRWZ7} and
\eqref{intBPT1}--\eqref{intBPT3} to compute the following quantities
\begin{eqnarray}
\label{inputt}
&&{\widetilde A}_{\ell},\ {\widetilde A}_{\ell}^{(1)},
\ {\widetilde B}_{\ell}^{(0)},\ {\widetilde B}_{\ell},\ {\widetilde Q}_{\ell},
\ {\widetilde F}_{\ell},\ {\widetilde D}_{\ell},\nonumber\\
&&{\widetilde T}_{(n)(n) \ell},\ {\widetilde T}_{(n)(\bar m)\ell},
\ {\widetilde T}_{(\bar m)(n\bar m) \ell}\ .
\end{eqnarray}
As a result, all of the above quantities are functions of $(t, r)$ only
and are evaluated at the discrete spacetime points $r_i,t_j$, where $r_i$
coincide with the radial gridpoints of the 2D code. The time levels
$t_j=j\Delta t\in[0,T]$, on the other hand, are chosen so that $T \gg
\tau$, where $\tau$ is the typical timescale for the problem ({\it e.g.,}
the torus oscillation timescale for the problem at hand and $T/\tau
\simeq 100$), and $\Delta t \ll \tau$ ({\it e.g.,} $\Delta t/\tau \simeq
10^{-2}-10^{-3}$). We note that in general the solution of the
hydrodynamical equations does not take place on equally spaced spacelike
hypersurfaces since the time step is automatically adjusted on the basis
of the dynamics of the matter source. As a result, the evaluation of the
functions~\eqref{inputt} on the equally spaced time levels $t_j$ requires
in general an interpolation process.

Once the interpolated timeseries for the functions (\ref{inputt}) have
been constructed at each gridpoint $r_i$, these are Fourier
transformed yielding the corresponding frequency-domain functions
\begin{eqnarray}
&&A_{\ell},\ A_{\ell}^{(1)},\ B_{\ell}^{(0)},\ B_{\ell},\  Q_{\ell},\ 
F_{\ell},\ D_{\ell},\nonumber\\ 
&&T_{(n)(n)\ell},\ T_{(n)(\bar m)\ell},\  T_{(\bar m)(n\bar m)\ell}
\label{inputw}
\end{eqnarray}
which are evaluated at the points $(r_i,\omega_j)$, with
$\omega_j=j\Delta\omega=j {2\pi}/{T}$. The functions~\eqref{inputw} are
then suitably combined as in equations (\ref{defsourceszrw1}),
(\ref{defsourceszrw2}) and (\ref{defsourcebpt}) to provide the compact
forms for the source terms in the frequency domain,
$S^{(\pm)}_{\ell}(r_i,\omega_j),\ S^{^{\rm BPT}}_{\ell}(r_i,\omega_j)$.

The next step consists of integrating the homogeneous equations
(\ref{ZRWhom}) to find the two independent solutions (\ref{eqZin}),
(\ref{eqZout}). The numerical solution is found using a $4^{\rm
  th}$-order Runge-Kutta integrator with adaptive stepsize and with
boundary conditions given by expressions~\eqref{eqZin} and
\eqref{eqZout}. Using the newly computed RWZ solutions $Z^{(\pm)~{\rm
    in/out}}_{\ell}(r_i,\omega_j)$ and the transformation
(\ref{chandra}), it is possible to also compute the two independent
solutions of the homogeneous BPT equation (\ref{BPThom}),
$\Psi_{\ell}^{\rm in/out}(r_i,\omega_j)$. Finally, the amplitude of the
emitted gravitational wave can be computed by numerically evaluating the
integrals (\ref{Hrwz}) and (\ref{Hbpt}), which then yield the energy
spectra (\ref{dEdw1bis}) and (\ref{dEdw2bis}).

With the exception of the handling of the solution of the relativistic
hydrodynamical equations, the procedure described so far for the
combination of the time and frequency-domain approaches is
straightforward and with minimal computational requirements. However,
great attention needs to be paid in to avoid unphysical results. In
particular, even when the timeseries extend over several tens of
dynamical timescales and the time sampling is also very high, the
calculation of the Fourier transforms can be inaccurate and this can be
problematic particularly at very large frequencies. Indeed, we have found
that the energy spectrum can become divergent at frequencies above a few
kHz even when the hydrodynamical evolution is carried out over $\sim 100$
dynamical timescales. 

The reason for this is most easily seen within the Green's function
approach; in this case, in fact, when $\omega$ increases above a few
kHz, the Wronskian of the homogeneous solutions tends to zero very
rapidly. At least in principle, this rapid decay should be compensated
by an equivalent decrease of the source function $S^{(\pm)}_{\ell}$ or
$S^{^{\rm BPT}}_{\ell}$ such that the wave amplitude would remain
finite.  In practice, however, this is not necessarily the case and if
the decay in the source functions is not sufficiently rapid, because
for instance their values at high frequencies are not sufficiently
accurate, this would inevitably lead to the high-frequency divergences
we have observed. Fortunately, a simple solution to this otherwise
serious problem is possible. In the limit of an infinite time
integration interval, in fact, it is possible to replace the Fourier
transform at a given frequency $\omega$ of a timeseries $h(t)$ with
the Fourier transform of its time derivative divided by $ \omega$,
{\it i.e.}
\begin{equation}
\label{ft_identity}
h(\omega) = \frac{{\rm i}}{\omega} \left(\frac{1}{2\pi} \int_{-\infty}^{\infty} 
\frac{d h(t)}{dt} e^{{\rm i} \omega t} dt \right) \ ,
\end{equation}
This identity, which is strictly true if $h(t) = 0$ for $t \to \pm
\infty$, can be exploited to compensate the loss of accuracy in the
Fourier transform. As a result, in our approach, and before Fourier
transforming the source terms $A_{\ell},\ A_{\ell}^{(1)}, \ldots$
in~\eqref{inputt}, we compute their first and second time derivatives,
evaluate the corresponding Fourier transform and finally divide the
result by $\omega^2$. This method effectively removes the divergence
appearing at large frequencies and has proven to work well also in the
case in which the timeseries is, in practice, finite.

%%%%%%%%%%%%%%%%%%%%%%%%%%%%%%%%%%%%%%%%%%%%%%%%%%%%%%%%%%%%%%%%%
\begin{table*}
\begin{center}
\caption{Main properties of the constant angular momentum toroidal
neutron star models used in the numerical calculations. From left to
right the columns report: the name of the model, the star-to-hole mass
ratio $M_{\rm t}/M$, the polytropic constant $\kappa$, the specific
angular momentum $\ell$ (normalized to $M$), the inner and outer radii of
the toroidal neutron star $r_\mathrm{in}$ and $r_\mathrm{out}$, the
radial position of cusp $r_\mathrm{cusp}$, the radial position of the
centre $r_\mathrm{centre}$ (all radii are in units of the gravitational
radius $r_\mathrm{g}$), and the orbital period at the centre of the torus
$t_\mathrm{orb}$, expressed in milliseconds.  The last two columns
indicate the density at the centre of the torus and the average density
of each model, respectively, both in cgs units. All of the models share
the same mass for the black hole, $M=2.5M_{\odot}$ and adiabatic index
$\gamma=4/3$.}
\label{tab1}
\vskip 0.5cm
\begin{tabular}{l|l|cc|cc|ccc|ccc}
\hline
Model   & $M_{\rm t}/M$    & $\kappa$      &
$\ell$  & $r_{\rm in}$     & $r_{\rm out}$ & $r_{\rm cusp}$
        & $r_{\rm centre}$ & $t_{\rm orb}$ & $\rho_{\rm centre}$
        & $\langle\rho\rangle$  \\
& & ${\rm (cgs)}$ & & & & & & ${\rm (ms)}$ & ${\rm (cgs)}$ &${\rm (cgs)}$ &\\

\hline
(c)   & 0.1    & 0.96${\times} 10^{14}$ &  3.8000  &4.576 &
        15.889 & 4.576  & 8.352 & 1.86 &1.14${\times} 10^{13}$ &4.73${\times}
        10^{11}$ \\
(e)   & 0.1    &   7.0${\times} 10^{13}$ & 3.7845 &4.646 &
        14.367 &4.646 & 8.165 & 1.80 &1.61${\times} 10^{13}$ &6.43${\times}
        10^{11}$  \\
(f)   & 0.1    &   1.0${\times} 10^{14}$ & 3.8022 &4.566 &
        16.122 &4.566 & 8.378 & 1.87 &1.10${\times} 10^{13}$ &4.48${\times}
        10^{11}$ \\
\hline
\end{tabular}
\end{center}
\end{table*}

%%%%%%%%%%%%%%%%%%%%%%%%%%%%%%%%%%%%%%%%%%%%%%%%%%%%%%%%%%%%%%%%%%%%%%%%%

%%%%%%%%%%%%%%%%%%%%%%%%%%%%%%%%%%%%%%%%%%%%%%%%%%%%%%%%%%%%%%%%%%%%%%%%%%% 
\section{A representative example: an oscillating
  high-density torus}
\label{results}
%%%%%%%%%%%%%%%%%%%%%%%%%%%%%%%%%%%%%%%%%%%%%%%%%%%%%%%%%%%%%%%%%%%%%%%%%%% 

The procedure discussed in the previous two Sections is totally generic
and could in principle be used to investigate the perturbations induced
on a Schwarzschild spacetime by an extended matter source when the
latter dynamics is simulated consistently in more than one spatial
dimension. However, as mentioned in the Introduction, we will here
concentrate on a proof of its effectiveness by considering the
perturbations induced by a non-selfgravitating torus orbiting and
oscillating around a Schwarzschild black hole. The latter will be
simulated with the 2D general relativistic code discussed in
Sect.~\ref{hydrosource}.

In order to to construct the background initial model for the torus,
which we subsequently perturb, we consider a perfect fluid described by
the stress-energy tensor (\ref{stress-tensor}) and in circular
non-geodesic motion with four-velocity $u^{\alpha} = (u^t,0,0,u^{\phi}) =
u^t(1,0,0,\Omega)$, where $\Omega = \Omega(r,\theta) \equiv u^{\phi}/u^t$
is the coordinate angular velocity as observed from infinity. Enforcing
the conditions of hydrostatic equilibrium and of axisymmetry simplifies
the hydrodynamics equations considerably and for a non-selfgravitating
fluid these reduce to Bernoulli-type equations
\begin{equation}
\label{bernoulli}
\frac{\nabla_i p}{e+p} = - \nabla_i W +
        \frac{\Omega \nabla_i l}{1- \Omega l} \ ,
\end{equation}
where $i=r,\theta$, $W \equiv \ln(u_t)$ and $l \equiv - u_{\phi}/u_{t}$
is the specific angular momentum.

%
%%%%%%%%%%%%%%%%%%%%%%  FIGURE  %%%%%%%%%%%%%%%%%%%%%%%%%%%%%%
\begin{figure}
\hskip -0.25cm
\includegraphics[angle=0,width=9.0cm]{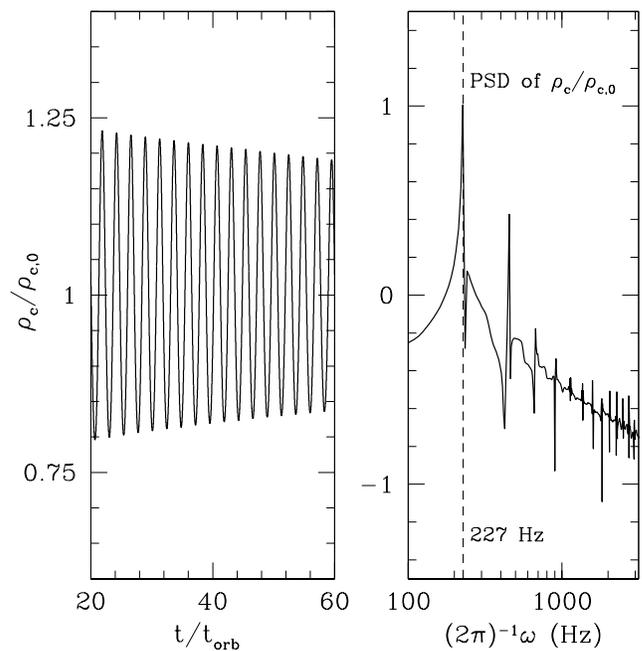}
\caption{Harmonic behaviour of an oscillating torus. Both panels refer to
  model (c) of Table~\ref{tab1} which was perturbed with a radial
  velocity having $\eta = 0.05$. The left panel shows the time variation
  of the maximum rest-mass density normalized to its initial value for a
  representative portion of the timeseries. The right panel shows the
  power spectral density (PSD) calculated over 100 orbital timescales,
  with the eigenfrequency of the $f$ mode indicated with a vertical
  dashed line at 227 Hz.}
\label{fig0}
\end{figure}
%%%%%%%%%%%%%%%%%%%%%%%%%%%%%%%%%%%%%%%%%%%%%%%%%%%%%%%%%%%%%%%%%%%%%%%%%%% 

The simplest solution to equations (\ref{bernoulli}) is the one with $l =
{\rm const.}$, since in this case the equipotential surfaces can be
computed directly through the metric coefficients and the value of the
specific angular momentum. Note that at any point in the $(r,\theta)$
plane, the potential $W$ can either be positive (indicating equipotential
surfaces that are open) or negative (indicating equipotential surfaces
that are closed). The case $W=0$ refers to that special equipotential
surface which is closed at infinity (see refs.~\cite{font:02,ZRF} for
details). local extrema on the equatorial plane of closed equipotential
surfaces mark the radial positions of the cusp, $r_{\rm{cusp}}$, and of
the ``centre'' of the torus, $r_{\rm c}$. At these radial positions the
specific angular momentum must be that of a Keplerian geodesic circular
orbit which can effectively be used to calculate the position of both the
centre and the cusp.

Stationary solutions with constant specific angular momentum are
particularly useful since in this case the angular velocity is fully
determined as $\Omega= l g_{tt} /g_{\phi \phi}$ and if a polytropic EOS
is used, the Bernoulli equations (\ref{bernoulli}) can be
integrated analytically to yield the rest-mass density (and pressure)
distribution inside the torus as
\begin{equation}
\label{density}
\rho(r,\theta) = \left\{\frac{\gamma-1}{\kappa \gamma}
        \left[\exp({W_\mathrm{in}-W})-1\right]\right\}^{1/(\gamma-1)} \ ,
\end{equation}
where $W_{\rm in} \equiv W(r_{\rm in},\pi/2)$. Clearly, different
configurations can be built depending on the value chosen for $l$, with
finite-extent tori resulting when $l_{\rm ms} < l < l_{\rm mb}$, with
$l_{\rm mb} = 4$, $l_{\rm ms} = 3\sqrt{3/2}$ being the the specific
angular momenta corresponding to orbits that are marginally bound or
marginally stable, respectively.

%
%%%%%%%%%%%%%%%%%%%%%%  FIGURE  %%%%%%%%%%%%%%%%%%%%%%%%%%%%%%
\begin{figure}
\includegraphics[angle=0,width=8.5cm]{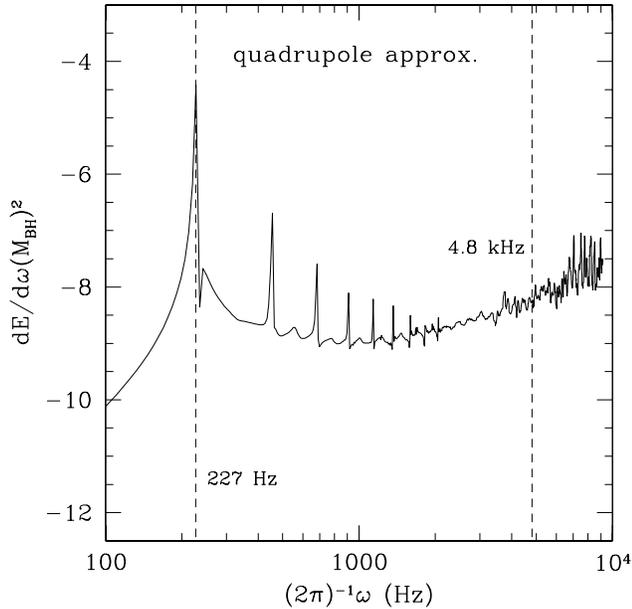}
\caption{Gravitational-wave energy spectrum produced by an oscillating,
  high-density torus orbiting around a 2.5 $M_{\odot}$ Schwarzschild
  black hole. The radiation has been computed for the $\ell=2$ mode using
  eq. (\ref{dEdw3}) within the Newtonian quadrupole approximation.  }
\label{fig1}
\end{figure}
%%%%%%%%%%%%%%%%%%%%%%%%%%%%%%%%%%%%%%%%%%%%%%%%%%%%%%%%%%%%%%%%%%%%%%%%%%% 

%%%%%%%%%%%%%%%%%%%%%%  FIGURE  %%%%%%%%%%%%%%%%%%%%%%%%%%%%%%
\begin{figure}
\includegraphics[angle=0,width=8.5cm]{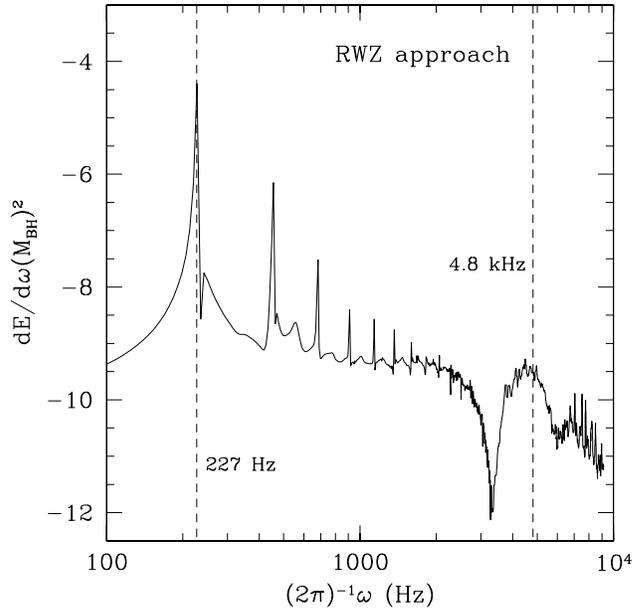}
\caption{ Same as Fig.~\ref{fig1} but  derived integrating  the RWZ equations 
in the frequency domain for $\ell=2$ and using eq. (\ref{dEdw1}).}
\label{fig2}
\end{figure}
%%%%%%%%%%%%%%%%%%%%%%%%%%%%%%%%%%%%%%%%%%%%%%%%%%%%%%%%%%%%%%%%%%%%%%%%%%% 

The configurations built in this way have a large quadrupole moment but
being stationary they do not produce any time-varying perturbation to the
background spacetime. Because of this we introduce parametrized
perturbations that would induce a small outflow through the cusp and
excite a quasi-harmonic behaviour in the hydrodynamical variables (the
interested reader is referred to refs~\cite{rmyz03,ryz03,zfrm05,mry04}
for a detailed discussion of the harmonic properties of this type of
oscillations). More specifically, we modify the stationary equilibrium
configuration with a small radial velocity which we have expressed in
terms of the radial inflow velocity characterising a relativistic
spherically symmetric accretion flow onto a Schwarzschild black hole,
{\it i.e.,} the Michel solution~\citep{michel:72}. Using $\eta$ to
parameterize the strength of the perturbation, we have specified the
initial radial (covariant) component of the three-velocity as
\begin{equation}
\label{vpert}
v_r = \eta (v_{r})_{_{\rm Michel}} \ .
\end{equation}

As a result of the introduction of the initial perturbation, the torus
acquires a linear momentum in the radial direction pushing it towards the
black hole. When this happens, the pressure gradients become stronger to
counteract the steeper gravitational potential experienced as the torus
moves inward, thus increasing the central density and eventually pushing
the torus back to its original position. The resulting oscillations are
essentially harmonic in the fundamental $f$-mode of oscillation but
other, higher-order $p$ modes are also excited and appear at frequencies
that are in a ratio of small integers~\cite{rmyz03,ryz03,zfrm05,mry04}.
This is summarised in the two panels of Fig.~\ref{fig0} which refer to
model (c) of Table~\ref{tab1} which was perturbed with a radial velocity
having $\eta = 0.05$ and evolved for several tens of the orbital
timescale $t_{orb}=1.86$ ms. The left panel shows as a function of time
the variation of the maximum rest-mass density ({\it i.e.,} the rest-mass
density at the centre of the torus $\rho_c$) normalized to its initial
value for a representative portion of the timeseries. The right panel, on
the other hand, shows as a function of the frequency $\nu \equiv
\omega/(2 \pi)$, the power spectral density (PSD) in arbitrary units of
the timeseries calculated over about 100 orbital timescales. Indicated
with a vertical dashed line at ${\bar \nu} = 227$ Hz is the
eigenfrequency of the fundamental ($f$) mode of oscillation and smaller
peaks at integer and semi-integer multiples of $\bar\nu$ are also visible
and are referred to the first overtones (see ref.~\cite{ryz03} for a
discussion of these modes).

\begin{figure}
\includegraphics[angle=0,width=8.5cm]{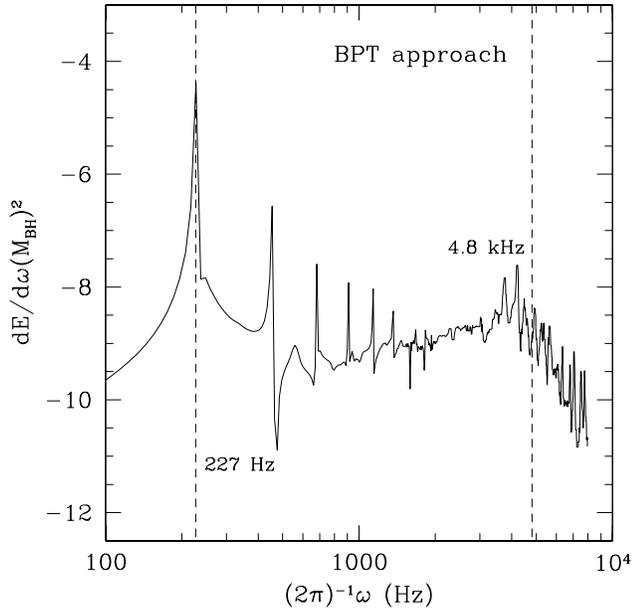}
\caption{Same as Fig.~\ref{fig1} but obtained integrating the BPT
  equation in the frequency domain for $\ell=2$ and using
  eq.~(\ref{dEdw2}).}
\label{fig3}
\end{figure}

As the simulation is carried out and the hydrodynamical equations are
solved along the lines discussed in Sect.~\ref{hydrosource}, the source
functions~\eqref{inputt} are calculated at the different radial
gridpoints and stored as distinct output. Once the dynamics has been
followed for a sufficient time-span comprising several tens orbital
timescales, the simulation in the time-domain is stopped and the values
of source functions~\eqref{inputt} throughout the simulated spacetime are
read-in at the different radial positions. This marks the begin of the
analysis in the frequency-domain, which first produces the source
functions~\eqref{inputw} and then proceeds to the solution of the
perturbation eqs.~\eqref{ZerRWeq} and~\eqref{bpt} along the lines
discussed in Sect.~\ref{RWZequation}. The results of the hybrid approach
are then summarised in Figs.~\ref{fig1}--\ref{fig3} which show the energy
spectrum of the emitted gravitational radiation as a function of the
frequency. 

The first figure, in particular, has been computed within the Newtonian
quadrupole formalism through expression~\eqref{dEdw3} and being a
faithful mirror of the hydrodynamical quantities, it shows a main peak at
227 Hz as well as the smaller peaks already discussed for
the right panel of Fig.~\ref{fig0}. Note that despite the use of the
procedure described at the end of Sect.~\ref{implementation}, the very
steep dependence on the frequency as $\omega^6$ of the energy spectrum
[{\it cf.} eq. \eqref{dEdw3}], produces an incorrect but finite growth at
high frequencies, where the accuracy in the calculation of the sources is
smaller. Of course, being rooted in a Newtonian approximation, the energy
spectrum in Fig.~\ref{fig0} shows no sign of a contribution coming from
the QNMs of the black hole that, for a mass of $M=2.5~M_\odot$ are
expected at $\nu =4.828~$kHz for the the fundamental $\ell=2$-QNM.

%%%%%%%%%%%%%%%%%%%%%%%%%%%%%%  FIGURE  %%%%%%%%%%%%%%%%%%%%%%%%%%%%%%
%%%%%%%%%%%%%%%%%%%%%%%%%%%%%%%%%%%%%%%%%%%%%%%%%%%%%%%%%%%%%%%%%%%%%%

%---> 

Fig.~\ref{fig2} shows instead the equivalent gravitational-wave energy
spectrum resulting from the hybrid approach employing the solution of the
RWZ equations [{\it cf.} eq. \eqref{dEdw1}] for $\ell=2$. A number of
interesting features should be noted. Firstly, at low frequencies the
spectra derived through the perturbative approach shows a rather good
agreement with the quadrupole spectrum given in Fig.~\ref{fig1}. This
confirms what already noted in~\cite{NFZD} and that, at least for this
type of sources, the quadrupole approach captures the most important
qualitative features of the energy spectrum.
Secondly, the spectrum is not diverging, nor
growing at large frequencies as a result of the use of the
transformation~\eqref{ft_identity}. Finally and most importantly, the
spectrum is not monotonically decreasing but shows a distinctive and
broad peak at $\nu \sim 4.4~$kHz, rather close to the position of the
expected fundamental $\ell=2$ QNM of the black hole and which we
interpret as the excitation of the QNMs of the black hole resulting from
the perturbations induced by the oscillating torus. Although its
contribution is energetically very small ({\it i.e.} less that
$2\times10^{-4}$ of the total emitted energy is produced at frequencies
larger than 2 kHz), the ability to isolate this peak is of great
importance to assess the validity of the hybrid approach and to prove its
effectiveness in modelling black hole spacetimes. The presence of such a
peak, in fact, is expected on the basis of simple considerations but was
{\it not detectable} in the complementary work of Nagar et
al.~\cite{NFZD} based on a time-domain perturbative approach despite the
corresponding PSD reached values was well below the ones at which our
black hole peak appears ({\it cf.}  Fig.\ref{fig2} with the left panel of
Fig.~1 in ref.~\cite{NFZD}).

Finally, we show in Fig.~\ref{fig3} the gravitational-wave energy
spectrum resulting from the hybrid approach employing the solution of the
BPT equation [{\it cf.} eq. \eqref{dEdw2}] for $\ell=2$. As expected,
much of what found for the RWZ equations continues to hold when the
hybrid approach is computed using the BPT equation, namely: very good
match with the quadrupole approximation at low frequency where most of
the energy is emitted and the appearance of a broad peak associated with
the black hole quasi-normal ringing.

%%%%%%%%%%%%%%%%%%%%%%%%%%%%%%%%%%%%%%%%%%%%%%%%%%%%%%%%%%%%%%%%%%%%%%%%%%% 
\section{Conclusions}
\label{conclusions}
%%%%%%%%%%%%%%%%%%%%%%%%%%%%%%%%%%%%%%%%%%%%%%%%%%%%%%%%%%%%%%%%%%%%%%%%%%% 

We have presented a new {\it hybrid} approach to study the perturbations
of a Schwarzschild black hole excited by extended matter sources in which
the solution in the {\it time-domain} of the relativistic hydrodynamical
equations in a multidimensional spacetime is coupled to the solution in
the {\it frequency-domain} of either the Regge-Wheeler/Zerilli equations
or of the Bardeen-Press-Teukolsky equations.

We believe that our hybrid approach may be preferrable over a pure
time-domain approach based on the RWZ equations for at least three
different reasons. Firstly, the Regge-Wheeler and the Zerilli equations
cannot be generalized to rotating black holes, whereas the BPT equation
admits such a generalization. Secondly, the BPT equation is
intrinsically unstable when integrated in the time domain but its
solution is regular computed in the frequency domain. Thirdly, for a
system undergoing small oscillations around an equilibrium state, a
frequency-domain approach is expected to be significantly more accurate
than one based in the time-domain, especially at high frequencies, where
the black hole contributions are expected to emerge. Within this
framework, therefore, the work presented here serves as a first step
towards the study of the gravitational-wave emission from an extended
source perturbing a rotating black hole.

As a test of its effectiveness and to provide a close comparison with
what done so far in perturbative approaches based in the time-domain, {\it
  i.e.,} ref.~\cite{NFZD}, we have considered as extended source to the
perturbation equations the oscillations of a high-density torus orbiting
around a Schwarzschild black hole. The dynamics of the torus has been
followed numerically using a 2D numerical code solving the general
relativistic hydrodynamics equations using HRSC schemes in spherical
polar coordinates.  

Overall, the results of our study show that the hybrid approach (either
with the RWZ equations or with the BPT one) works remarkably well,
producing an energy spectrum of the emitted gravitational radiation which
contains both the basic quadrupolar emission of the torus as well as the
excited emission of the black hole.  The latter feature was completely
washed out in the complementary time-domain approach of ref.~\cite{NFZD}.

Being totally generic, the formalism and methodology presented here is
could be used under more general conditions and in particular also with
sources coming from 3D numerical relativity codes as long as an accurate
evolution of the matter sources is possible. Future work in this line of
research will therefore comprise both the extension of the present
results to 3D codes, as well as the generalization of the perturbative
formalism to the Teukolsky equation in order to model the gravitational
emission of rotating black holes when perturbed by extended sources.

\acknowledgments 

It is a pleasure to thank J.~A. Font, A.~Nagar, J.~A.Pons and O. Zanotti
for useful comments and discussions. Support to this research comes also
through the SFB-TR7 ``Gravitationswellenastronomie'' of the DFG and the
Italian INFN, {\it Iniziativa Specifica} OG51.  The computations were
performed on the Beowulf Cluster for numerical relativity {\it
  ``Albert''} at the University of Parma.

%%%%%%%%%%%%%%%%%%%%%%%%%%%%%%%%%%%%%%%%%%%%%%%%%%%%%%%%%%%%%%%%%%%%%%%%%
\appendix 
\section{Generic source terms}
\label{sources} 
%%%%%%%%%%%%%%%%%%%%%%%%%%%%%%%%%%%%%%%%%%%%%%%%%%%%%%%%%%%%%%%%%%%%%%%%%%% 

%--------------------------------------------------------------------------
\subsection{Source for the RWZ equations}
\label{sources_rwz}
%--------------------------------------------------------------------------

The source of the RWZ equations (\ref{ZerRWeq}),
$S_{\ell}^{(\pm)}(r,\omega)$, can be expressed in terms of the components
of the stress-energy tensor $T_{\mu\nu}(t,r,\theta,\phi)$ as follows. We
first decompose $T_{\mu\nu}$ in tensor spherical harmonics for the 
the relevant components
\begin{eqnarray}
T_{tr}(r,t)&=&\frac{{\rm i}}{\sqrt 2}\sum_{\ell m}{\widetilde A}^{(1)}_{\ell m}Y^{\ell  m}
	\ ,\\%\nonumber 
T_{rr}(r,t)&=&\sum_{\ell m}{\widetilde A}_{\ell m}Y^{\ell m}
	\ ,\\%\nonumber 
T_{ta}(r,t)&=&\frac{{\rm i} r}{\sqrt{2\Lambda^{\prime}}}
  \sum_{\ell m}\left({\widetilde B}^{(0)}_{\ell m}Y^{\ell m}_{\ \ \ ,a} 
	-{\widetilde Q}^{(0)}_{\ell m}S^{\ell m}_a\right)
	\ ,\\%\nonumber
T_{ra}(r,t)&=&\frac{r}{\sqrt{2\Lambda^{\prime}}}\sum_{\ell m}
	\left({\widetilde B}_{\ell m}Y^{\ell m}_{\ \ \ ,a} 
	-{\widetilde Q}_{\ell m}S^{\ell m}_a\right)
	\ ,\\%\nonumber
T_{ab}(r,t)&=&\frac{r^2}{\sqrt 2}\sum_{\ell m}
	\left({\widetilde G}_{\ell m}Y^{\ell m}\gamma_{ab}+ 
	\right. \nonumber \\
	&& \hskip 2.0cm 
	\left.\frac{{\widetilde F}_{\ell m}Z_{ab}^{\ell m}- 
	{\rm i} {\widetilde D}_{\ell
      m}S^{\ell m}_{ab}}{\sqrt{2\Lambda\Lambda^{\prime}}} 
  \right)\ ,\nonumber\\
\label{expansion}
\end{eqnarray}
where $a=\theta,\phi$, $\gamma_{ab}\equiv{\rm diag}(1,\sin^2\theta)$ is the
metric tensor of the sphere, $Y^{\ell m}$ are the scalar spherical
harmonics, $Y^{\ell m}_{\ \ \ , a} \equiv \partial Y^{\ell m}/\partial
x^a$ and and ${\Lambda} \equiv (\ell-1)(\ell+2)/2$,
${\Lambda^{\prime}} \equiv \ell(\ell+1)$. The matrices
\begin{eqnarray}
S^{\ell m}_a&=&\left(-Y^{\ell m}_{\ \ \ ,\phi}/ {\sin\theta},
\sin\theta Y^{\ell m}_{\ \ \ ,\theta}\right)\ ,\\%\nonumber
Z^{\ell m}_{ab}&=&\left(\begin{array}{cc} W^{\ell m} & X^{\ell m} \ ,\\%\nonumber 
X^{\ell m} & -\sin^2\theta W^{\ell m} \\\end{array}\right)\ ,\\%\nonumber
S^{\ell m}_{ab}&=&\left(\begin{array}{cc} 
-{X^{\ell m}}/{\sin\theta} & 
\sin\theta W^{\ell m} \\ \sin\theta W^{\ell m} & \sin\theta X^{\ell m} \\ 
\end{array}\right)\ ,
\end{eqnarray}
with
\begin{eqnarray}
W^{\ell m}&=&Y^{\ell m}_{,\theta\theta}-\cot\theta Y^{\ell m}_{\ \ \ ,\theta}
+\frac{m^2}{\sin^2\theta}Y^{\ell m}\ ,\\%\nonumber
X^{\ell m}&=&2{\rm i} m\left(Y^{\ell m}_{\ \ \ ,\theta}-\cot\theta Y^{\ell m}\right)\ ,
\end{eqnarray}
are vector and tensor spherical harmonics, and 
collect all of the angular dependence.

Next, using the orthogonality properties of $Y^{\ell m},\ S_a^{\ell m},\
  S_{ab}^{\ell m},\ Z_{ab}^{\ell m}$ and the axial symmetry of the
  stress-energy tensor it is possible to show that only the $m=0$
  harmonics contribute to the expansion (\ref{expansion}) and that
\begin{eqnarray}
\label{intRWZ1}
&&\hskip -0.75cm {\widetilde A}_{\ell}=2\pi\int d\theta\sin\theta Y^{\ell 0}T_{rr}\ ,\\%\nonumber
&&\hskip -0.75cm {\widetilde A}_{\ell}^{(1)}=-{\rm i}\sqrt{2}2\pi
\int d\theta\sin\theta Y^{\ell 0}T_{tr}\ ,\\%\nonumber
&&\hskip -0.75cm {\widetilde B}_{\ell}^{(0)}=-{\rm i}\lambda_12\pi
\int d\theta\sin\theta T_{t\theta}Y^{\ell 0}_{\ \ \ ,\theta}\ ,\\%\nonumber
&&\hskip -0.75cm {\widetilde B}_{\ell}=\lambda_12\pi
\int d\theta\sin\theta T_{r\theta}Y^{\ell 0}_{\ \ \ ,\theta}\ ,\\%\nonumber
&&\hskip -0.75cm {\widetilde Q}_{\ell}=\lambda_12\pi
\int d\theta\sin\theta\frac{1}{\sin\theta}T_{r\phi}Y^{\ell 0}_{\ \ \ ,\theta}
\ ,\\%\nonumber
&&\hskip -0.75cm {\widetilde F}_{\ell}=\lambda_22\pi\int d\theta\sin\theta
\left(T_{\theta\theta}-\frac{1}{\sin^2\theta}T_{\phi\phi}\right)W^{\ell 0}
\ ,\\%\nonumber
\label{intRWZ7}
&&\hskip -0.75cm {\widetilde D}_{\ell}=-{\rm i}\lambda_22\pi
\int d\theta\sin\theta\left(-\frac{2}{\sin\theta}T_{\theta\phi}\right)
W^{\ell 0}\ ,
\end{eqnarray}
where 
\begin{eqnarray}
\lambda_1&\equiv&\frac{\sqrt{2}}{r\sqrt{{\Lambda^{\prime}}}}\ ,\\%\nonumber
\lambda_2&\equiv&\frac{1}{r^2\sqrt{4{\Lambda}{\Lambda^{\prime}}}}\ .
\end{eqnarray}

In terms of the Fourier transform of these quantities, the source terms
$S_{\ell}^{(\pm)}(r,\omega)$ are
\begin{eqnarray}
\label{defsourceszrw1}
S^{(+)}(r,\omega)&=&\alpha_0 A^{(1)}+\alpha_1 A^{(1)}_{,r}
+\beta  A+\gamma  B+\delta F \nonumber\\
&&+\epsilon_2 B^{(0)}_{,rr}+\epsilon_1 B^{(0)}_{,r}
+\epsilon_0 B^{(0)}
\ ,\\%\nonumber 
\label{defsourceszrw2}
S^{(-)}(r,\omega)&=&\sigma_1\left(-g_{00} D\right)_{,r}
+\chi Q\ ,
\end{eqnarray}
with 
\begin{eqnarray}
\alpha_0&=&4\pi\sqrt{2}\frac{M(r-2M)((\Lambda +3)r-3M)}{r(\Lambda
r+3M)^2\omega} \ ,\\%\nonumber
\alpha_1&=&2\pi\sqrt{2}\frac{(r-2M)^2}{\omega(\Lambda r+3M)}\
,\\%\nonumber 
\beta&=&4\pi\frac{(r-2M)^2}{\Lambda r+3M}\ ,\\%\nonumber
\gamma&=&4\pi\sqrt{\frac{2}{\Lambda^{\prime}}}\frac{(r-2M)^2}{\Lambda
r+3M}\ ,\\%\nonumber
\delta&=&-\frac{8\pi}{\sqrt{{\Lambda}{\Lambda^{\prime}}}}(r-2M)\
,\\%\nonumber 
\epsilon_2&=&-4\pi\sqrt{\frac{2}{\Lambda^{\prime}}}
\frac{r(r-2M)^2}{\omega(\Lambda r+3M)}\ ,\\\nonumber
\epsilon_1&=&-4\pi\sqrt{\frac{2}{\Lambda^{\prime}}}\times\nonumber\\
&&\hskip 0.5cm \frac{(r-2M)(3\Lambda r^2+15Mr-4M\Lambda
r-24M^2)}{\omega(\Lambda r+3M)^2}\ ,\nonumber \\ \\
\epsilon_0&=&4\pi\sqrt{\frac{2}{\Lambda^{\prime}}}
\frac{1}{\omega}\times\nonumber\\
&& \hskip 0.5cm \left[\frac{(r-2M)(\Lambda ^2r^2-3M\Lambda
r-12Mr+12M^2)}{r(\Lambda r+3M)^2}\right. \nonumber \\ 
&& \hskip 0.5cm
\left.-\frac{\omega^2 r^3}{\Lambda r+3M}\right]\ ,\\%\nonumber
\chi&=&-8\pi {\rm i}\sqrt{\frac{2}{\Lambda^{\prime}}}
\frac{(r-2M)^2}{r^2}\ ,\\%\nonumber 
\sigma_1&=&{\rm
i}\frac{8\pi }{\sqrt{{\Lambda}{\Lambda^{\prime}}}}(r-2M)\ .
\end{eqnarray}

%--------------------------------------------------------------------------
\subsection{Source for the BPT equation}
\label{sources_bpt}
%--------------------------------------------------------------------------

The source of the BPT equation (\ref{bpt}), $S_{\ell}^{^{\rm
BPT}}(r,\omega)$, can be expressed in terms of the components of the
stress-energy tensor as follows (see~\cite{Teu,SN})
\begin{eqnarray}
&&S_{\ell}^{^{\rm BPT}}(r,\omega)=-\sqrt{2{\Lambda}{\Lambda^{\prime}}}r^4
	T_{(n)(n)\  \ell} \nonumber \\
&& \hskip 1.5cm
	-\sqrt{2{\Lambda^{\prime}}}\Delta{D_+}\left[\frac{r^5}{\Delta}
	T_{(n)(\bar m)\  \ell}\right] \nonumber \\
&& \hskip 1.5cm
	-\frac{\Delta}{2r}{D_+} \left[\frac{r^6}{\Delta} D_+\left(r
	T_{(\bar m) (\bar m)\  \ell}\right)\right]\ ,
\label{defsourcebpt}
\end{eqnarray}
where ${D}_+\equiv {d}/{dr_*}+i\omega$, 
and $ T_{(n)(n)\  \ell},T_{(n)(\bar m)\  \ell},T_{(\bar m) (\bar m)\  \ell}$ 
are the Fourier transform of the following quantities
\begin{eqnarray}
\label{intBPT1}
&&\hskip -0.5cm {\widetilde T}_{(n)(n)\  \ell}\equiv 
	\int d\Omega~_0S_{\ell 0}(\theta)
        T^{\mu\nu}n_{\mu}n_{\nu}\ ,\\%\nonumber 
&&\hskip -0.5cm {\widetilde T}_{(n)(\bar m)\ 
          \ell}\equiv \int d\Omega~_{-1}S_{\ell 0}(\theta)
        T^{\mu\nu}n_{\mu}\bar m_{\nu}\ ,\\%\nonumber 
&&\hskip -0.5cm {\widetilde T}_{(\bar m)(\bar m)\ 
          \ell}\equiv \int d\Omega~_{-2}S_{\ell 0}(\theta) T^{\mu\nu}\bar
        m_{\mu}\bar m_{\nu}\ , 
\label{intBPT3}
\end{eqnarray}
where $d\Omega = \sin\theta d\theta d\phi$ is the solid angle and the
one-forms $n_\mu$, $\bar m_\mu$ are defined as 
\begin{eqnarray}
n_{\mu}&\equiv&\frac{1}{2}\left(1-\frac{2M}{r},1,0,0\right)\ ,\\%\nonumber
\bar m_{\mu}&\equiv&-\frac{1}{\sqrt{2}}\left(0,0,r,-{\rm i} r\sin\theta\right)\ .
\end{eqnarray}
Here, $_{-s}S_{\ell m}(\theta,\phi)$ are the spin-weighted spherical
harmonics and we have considered only those with $m=0$ because of the
underlying axisymmetry in the perturbations.

\end{document}